\documentstyle[prl,aps,multicol]{revtex}
\input{epsf}
\begin{document}
\draft
\title{Quantum Fractal Eigenstates}                      

\author{Giulio Casati$^{(a,b,c)}$, Giulio Maspero$^{(a,b,c)}$ and 
Dima L. Shepelyansky$^{(d,+,*)}$}
\address{$^{(a)}$Universit\`a di Milano, sede di Como, Via Lucini 3,
22100 Como, Italy}
\address{$^{(b)}$Istituto Nazionale di Fisica della Materia, 
Unit\`a di Milano, Via Celoria 16, 20133 Milano, Italy}
\address{$^{(c)}$Istituto Nazionale di Fisica Nucleare, Sezione di Milano,
Via Celoria 16, 20133 Milano, Italy} 
\address {$^{(d)}$ Isaac Newton Institute, 20 Clarkson Rd, Cambridge,
 CB3 0EH, U.K. }

\date{\today}
\maketitle
\begin{abstract}
We study quantum chaos in open dynamical systems and show that 
it is characterized by
quantum fractal eigenstates located on the underlying classical
strange repeller. The states with longest life times
typically reveal a scars structure  on the classical
fractal set.
\end{abstract}

\begin{multicols}{2}
\narrowtext

At present, the structure of phase space for generic
 classical Hamiltonian systems is 
well understood both on qualitative and quantitative levels.
With the increase of perturbation's parameter the invariant
KAM curves are destroyed by the perturbation and are replaced 
by chaotic regions. These regions grow and become interconnected
over the whole phase space. In this situation the phase space exhibits
a hierarchical structure of mixed, chaotic and integrable components which
continues on smaller and smaller scales of the phase space. At larger
perturbations the measure of stability islands decreases until 
they become negligible for sufficiently strong perturbations. 
This is a general scenario for the emergence of chaos in 
classical Hamiltonian systems \cite{Licht}. The well known model
in which such scenario has been studied in detail is the Chirikov
standard map \cite{Chirikov}.

The investigations of the corresponding quantum systems show that
the structure of eigenstates is closely related to the properties
of classical phase space. In the integrable regime, eigenstates are located on 
the invariant KAM curves while in the chaotic regime they spread
over the whole chaotic component in agreement with the Shnirelman theorem
\cite{Shnir}.
The structure of eigenstates can be seen in a pictorial way with the help 
of Wigner function and Husimi distribution \cite{Husimi}.
This representation allows to see graphically the qualitative change
of eigenstates during the transition from integrability to chaos. 
In the chaotic regime they also allow to see scarred eigenfunctions in
which the probability is concentrated near short unstable  periodic orbits
\cite{Heller,Agam}.
However in the chaotic regime, the majority of eigenfunctions are ergodic
on the energy surface and the energy level statistics is well described
by the Random Matrix Theory \cite{Bohigas}.
Here we assume that the value of the Planck constant is sufficiently
small and the effect of dynamical localization does 
not alter the properties of eigenstates \cite{ChirHouch}.

As a result, in absence of localization, 
 the structure of quantum eigenstates in the regime 
of Hamiltonian chaos is now well understood 
and many concrete examples have been studied numerically 
\cite{Heller,Agam,Bohigas,ChirHouch} and experimentally \cite{Bill}.
Conversely, the chaotic objects with fractal structure,
which appear in dissipative classical
dynamics such as strange attractors and repellers, were not studied in
quantum mechanics.
The main problem is that in the quantum case the dissipation is always 
accompanied by noise and generally one should study the density
matrix \cite{Haake,Graham}.
In this way the problem becomes much more 
complicated than the Hamiltonian case and the fractal
structure of strange attractors/repellers had  never been seen in quantum 
eigenstates.

In this paper we study an open chaotic system in which absorption
leads to the appearance of a fractal set 
in the classical phase space (strange repeller) . The quantum dynamics 
of the model is governed by a nonunitary evolution operator. 
Such unitary breaking,  due to absorption, was widely used in
nuclear and mesoscopic physics  to describe  coupling 
with continuum and massive leads via open channels\cite{Weid}. 

The underlying classical fractal set should affect the quantum 
dynamics and find its manifestations in the structure of eigenstates.
Indeed it is natural to expect that long living
eigenstates will be associated with the above strange set on which 
classical orbits live forever. Therefore the eigenstates associated
with this set should strongly influence the scattering
process, relaxation and ionization into continuum.

To investigate  the quantum fractal eigenstates we choose the kicked rotator
model with absorbing boundary conditions which was introduced in \cite{Borg}.
The quantum dynamics is described by the evolution matrix:
\begin{eqnarray} 
\label{qmap}
\bar{\psi} = \hat{U} \psi = \hat{P} e^{-iT\hat{n}^2/4} e^{-ik
\cos{\hat{\theta}}}
 e^{-iT\hat{n}^2/4} \psi,
\end{eqnarray}
where $\hat{n}=-i {\partial / { \partial \theta}}$,
 $\hbar=1$  and 
the operator $\hat{P}$ projects the wave function to the states in the
interval $[ -N/2, N/2 ]$.
The quasiclassical limit corresponds to $k \rightarrow \infty$, 
$T \rightarrow 0$ with
the  chaos parameter $K=kT=const$.

The classical  dynamics is described by the Chirikov standard map 
\cite{Chirikov}:
\begin{eqnarray} \label{cmap}
\bar{n} = n + k \sin{ \left[ \theta + { T n \over 2} \right]},
\bar{\theta} = \theta + {T \over 2} (n+\bar{n}).
\end{eqnarray}
In this model all  trajectories 
(and quantum  probabilities) leaving the interval $[ -N/2, N/2 ]$
are absorbed and never return back.
For the classical map 
the ionization time required for a trajectory to reach diffusively
the absorbing boundary is $t_c \sim t_D = N^2 /D$ where $D \approx k^2 /2$ 
is the diffusion rate for $K \gg 1$. The independence of 
$t_c$ on $N$ requires $N/k=const$.

In our study we fixed $N/k=4$ and $K=7$ so that the
phase space is completely chaotic with no visible islands. As
a result the classical probability $P(t)$ to stay inside the sample 
$[ -N/2, N/2 ]$ decays exponentially for $t > t_D$: $P(t) \sim
\exp{\left(-\gamma_ct\right)}$ with $\gamma_c=1/t_c=0.10188$.
The quantum probability follows closely
the classical one up to the quantum relaxation
time scale $t_q \sim \sqrt{t_c/\Delta}$ \cite{second} 
where $\Delta = 1/N$ is the levels spacing.

In the classical case, the orbits which are never ionized and which stay
forever inside the sample form a fractal set (strange repeller).
To obtain this set we study the evolution of $M=2.2 \cdot 10^9$ classical
orbits up to $t=100 $ map iterations. Initially the orbits are 
homogeneously distributed in the phase plane.
An example of a set of points which are never ionized and form a 
fractal is shown in Fig.1. The fractal nature is
demonstrated by magnification of a small part of the phase space which clearly 
shows a structure typical of strange attractors/repellers \cite{Licht,sa}.
Similarly to the problem of diffusion in the Lorentz gas \cite{Gaspard}, the
information dimension $d_1$ of this set can be expressed  via the
Lyapunov exponent $\Lambda \approx \ln{K/2}$ and
the probability decay rate $\gamma_c \sim D/N^2$: $d_1 =2-\gamma_c/\Lambda$.
In our case with $\gamma_c \approx 0.1$, $\Lambda \approx 1.25$ this gives 
$d_1 \approx 1.92$.

In the quantum system, due to absorption at the
boundary, all eigenvalues of $ \hat{U}$ move inside the unitary circle
and can be written as $\lambda = \exp{(-i\epsilon)}= \exp{(-iE-\Gamma/2)}$.
The imaginary part of $\epsilon$ determines the decay rate $\Gamma$ of an
eigenstate.
Due to the symmetry of the $U$-matrix the eigenstates are symmetric or
antisymmetric in $n$ and in order to study the statistics of $\Gamma$'s 
we restricted ourselves to investigation of 
symmetric states.
The general structure of the distribution of $\Gamma$'s had been
found in \cite{Borg}. This distribution $dW /{d\Gamma}$ has a gap for small
values of $\Gamma$ ($\Gamma < \gamma_c$) while for $\Gamma > \gamma_c$ it
drops according to a power law, ${dW / {d\Gamma}}
\sim \Gamma^{-{3\over2}}$ for $t_c >> 1$ \cite{Borg}.
A typical example for $N=12001$ is shown in Fig.2
and confirms the above global structure of the distribution.
With the increase of the matrix size $N$, the minimal value of 
$\Gamma=\Gamma_{min}$ converges to the classical value $\gamma_c$ as 
it is shown in the inset of Fig.2. The fit of numerical data  gives 
$\Delta \Gamma = \gamma_c - \Gamma_{min} 
\approx 6.2 \gamma_c g^{-\alpha}$ with $\alpha=0.507 \pm 0.037$.
Theoretically we expect that, due to fluctuations,
$\Delta \Gamma$ should be
of the order of the distance $\delta$ between the eigenvalues of $\hat{U}$
in the complex plane. Since most of the  $N$ eigenvalues $\epsilon$ are  
distributed inside the ring of width $\gamma_c$, then
$\delta \sim \left( \gamma_c /N\right)^{1/2}  \sim
1/t_q$.  This gives $\Delta \Gamma \sim \delta \sim 
1/t_q$. 
Therefore, after the fitting of the data in Fig.2, we have
\begin{eqnarray} \label{rel1}
\Delta \Gamma = \gamma_c-\Gamma_{min} \approx 6.2 \gamma_c/\sqrt{g}.
\end{eqnarray}
where $g=\gamma_c/\Delta$ is the effective conductance of the system.

To analyze the structure of eigenfunctions in the phase space we used the
Husimi function obtained from the Wigner function smoothed in the
intervals $\Delta n $ and $\Delta \theta$ $(\Delta n \Delta \theta =1/2)$
(see \cite{Husimi}).
The ratio $s=\Delta \theta /\Delta n $ was fixed in a way to have
optimal resolution. In Fig.1 it was equal to $s=0.0015 (a);
6.25 \times 10^{-4} (c); 2.25 \times 10^{-4} (e)$ and in Fig.3
it was $s=0.0015 (a); 0.16 (b,c)$. The Husimi functions were constructed 
from antisymmetric eigenstates since they had the minimal values of $\Gamma$. 
The comparison between the density distributions in the phase space
for the classical and the quantum case can be seen in Fig.1. The quantum
Husimi function for the eigenstates with $\Gamma \approx \gamma_c$ reproduces
very well the fractal structure of the classical strange repeller
on very small scales. It shows close agreement between
the classical and quantum data for large $N$ corresponding to
a small effective Planck constant $\hbar_{eff}$.
Of course, on very small scales comparable with the minimal
quantum cell, the quantum density becomes smooth.
However, on  scales larger than this cell, the fractal structure
is obvious: we call such states {\it quantum fractal eigenstates}.

It is natural to expect some analogous of Shnirelman theorem \cite{Shnir}
for these quantum fractal eigenstates, so that in the quasiclassical
limit they will be distributed over the classical fractal set according 
to the classical measure. The case of Fig.1 for the states with 
$\Gamma \approx \gamma_c$ confirm such expectations.
However, the situation for the states with $\Gamma = \Gamma_{min}$
is rather different (Fig.3). Indeed, generally, 
we observe there the appearance of scars on the underlying classical strange
repeller, the silhouette of which, in spite of scars, is still clearly seen.
The strength of scars grows with decreasing $N$ (increasing $\hbar_{eff}$).
However, even for the largest $N=59049$, the probability distributions 
$f(n)$ projected on the $n$-axis demonstrate an evident difference
between the classical and quantum cases (Fig.4).
We qualitatively understand this phenomenon in the following way: 
the quantum state with the minimal value of $\Gamma$ should stay away as far 
as possible from the absorbing  boundaries so that 
quantum interference should redistribute probability on the classical 
set and will lead to some scarring around classical trajectories
which correspond to short unstable periodic orbits located near the center.
Indeed Fig.4
shows a pronounced peak near period two orbits. There is also some 
correlation between the density of long periodic orbits (Fig.3d) and quantum
distributions with not very large $N$ (Fig.3b,c).
However, at present, we cannot propose any quantitative
explanations of the scarred eigenstate structure.

The complex eigenvalues of the evolution operator $U$ can be considered as
some poles of the scattering matrix. Therefore, we can expect that similar
quantum fractal eigenstates will appear in the problems of chaotic 
scattering in the quasiclassical regime. One of the possible models
to study such effects is the three discs problem where the gap 
in the $\Gamma$-rates has been discussed in \cite{Gaspard1}.
Such quantum fractal eigenstates can be also studied
in experiments with chaotic light in micrometer-size droplets \cite{Stone}
where the classical dynamics is governed by a map analogous to the map (2).
In conclusion, we have demonstrated that the quantum eigenstates
can form a quantum strange repeller. 
We conjecture that quantum strange attractors, once identified, should have a 
similar structure.

We thank B.Georgeot for the discussions about the Shnirelman theorem for 
nonunitary operators and R.Ketzmerick for useful advises
in Lanczos approach to the kicked rotator.

\begin{figure}
\caption{Left column: Husimi function of the quantum fractal eigenstate
for $N=59049$ and $\Gamma = 0.1080 \approx \gamma_c$.
Right column: classical strange repeller obtained,
from initially homogeneously distributed orbits,
after 100 iterations of map (2).
The color varies proportionally to the
density: from blue for zero density to bright red for maximal
density (the color scale is the same for
 classical and  quantum cases on each magnification level). 
The size of the phase region 
$(\theta, n)$ is: $0 \leq \theta \leq 2 \pi$,
$-N/2 \leq n \leq N/2$ for (a) and (b); $0 \leq \theta \leq 2 \pi/3$,
$-N/6 \leq n \leq N/6$ for (c) and (d); $0 \leq \theta \leq 2 \pi/9$,
$-N/18 \leq n \leq N/18$ for (e) and (f).
}
\end{figure}

\begin{figure}
\caption{ 
Distribution $d W/d \Gamma$
as a function of  $\Gamma$ for $N=12001$.
The arrow marks the value of $\gamma_c$.
The inset shows  $\Delta \Gamma = \gamma_c - \Gamma_{min}$
vs. $N$ in log-log scale. The straight line shows the dependence given by
(3). To guide the eye, the numerical data (points) are connected by lines.
}
\end{figure}

\begin{figure}
\caption{
Husimi function for quantum eigenstates with
minimal $\Gamma=\Gamma_{min}$ for: (a) $N=59049$;
(b) $N=729$ and (c) $N=243$. The case (d) shows the classical initial density
of orbits which survive after  60 iterations forward and back
in time. The maximum of density (bright red) from (a) to (d)  is fixed 
by case (a).
}
\end{figure}

\begin{figure}
\caption{Probability distribution $f(n)$ over the unperturbed basis $n$
for quantum eigenstates of Fig.1a (dotted curve),
Fig.3a (dashed curve). The full curve is the classical distribution 
of Fig.1b.
Crosses mark periodic orbits 
with period one at (0,0) and period two at ($n=\pm \pi/T$ for 
$\theta=\pi/2, 3\pi/2$). The asymmetry of $f(n)$ in $n$ for the classical case
is due to round off computer errors.
}
\end{figure}
\end{multicols}

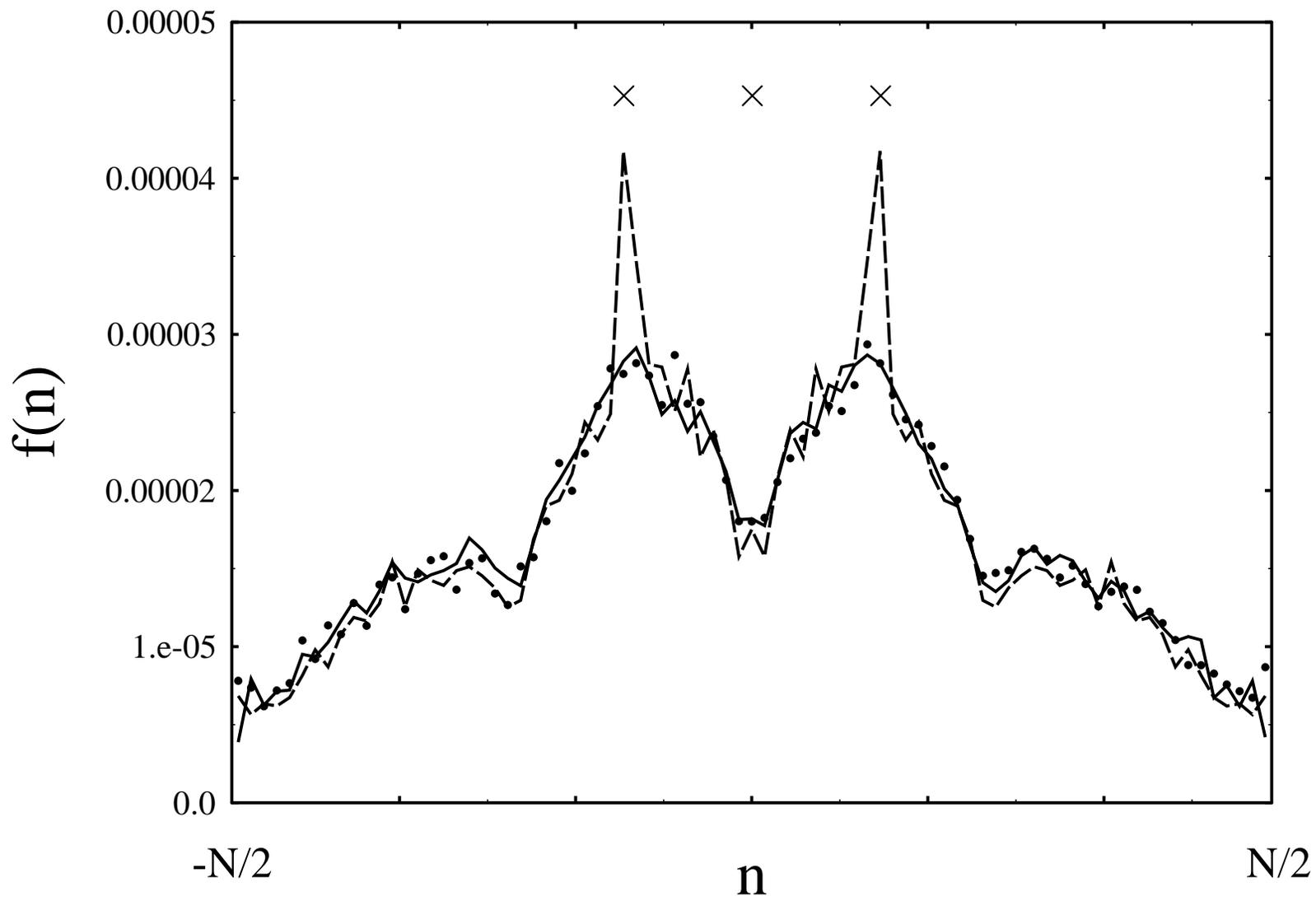
\end{document}